# Sculpturing the Electron Wave Function

Roy Shiloh, Yossi Lereah, Yigal Lilach and Ady Arie

*Department of Physical Electronics, Fleischman Faculty of Engineering, Tel Aviv University, Tel Aviv 6997801, Israel*

*Coherent electrons such as those in electron microscopes, exhibit wave phenomena and may be described by the paraxial wave equation[1]. In analogy to light-waves[2,3], governed by the same equation, these electrons share many of the fundamental traits and dynamics of photons.*

*Today, spatial manipulation of electron beams is achieved mainly using electrostatic and magnetic fields. Other demonstrations include simple phase-plates[4] and holographic masks based on binary diffraction gratings[5-8]. Altering the spatial profile of the beam may be proven useful in many fields incorporating phase microscopy[9,10], electron holography[11-14], and electron-matter interactions[15]. These methods, however, are fundamentally limited due to energy distribution to undesired diffraction orders as well as by their binary construction.*

*Here we present a new method in electron-optics for arbitrarily shaping of electron beams, by precisely controlling an engineered pattern of thicknesses on a thin-membrane, thereby molding the spatial phase of the electron wavefront.*

*Aided by the past decade's monumental leap in nano-fabrication technology and armed with light-optic's vast experience and knowledge, one may now spatially manipulate an electron beam's phase in much the same way light waves are shaped simply by passing them through glass elements such as refractive and diffractive lenses. We show examples of binary and continuous phase-plates and demonstrate the ability to generate arbitrary shapes of the electron wave function using a holographic phase-mask.*

*Our results make evident that the light-optics concept of multilevel diffractive or refractive phase elements may now be harnessed in full employing nothing else but nano-fabrication machinery. This opens exciting new possibilities for microscopic studies of materials using shaped electron beams and enables electron beam lithography without the need to move the electron beam or the sample, as well as high resolution inspection of electronic chips by structured electron illumination.*

With the advance of nano-fabrication technology, new possibilities have opened for fundamental research in electron optics. Nano-scale wrought thin films, using a focused ion-beam (FIB) for example, can be fashioned as phase-masks to create holograms that may be utilized in material science and in fabrication of microelectronic circuits.

A number of groups have proposed analogies between free-electron and light optics. Specifically, it has been suggested that under certain approximations feasible in a standard transmission electron microscope (TEM), the Klein-Gordon equation describing the dynamics of free electrons may be replaced by the paraxial Helmholtz wave propagation equation, which is used in light optics[1]. One of the first realizations in this direction was already made in 1998 - a Fresnel lens in $AlF_3$ film was fabricated by electron beam nano-lithography[16]. More recently, Uchida and Tonomura measured vortex electron beams[17], having a helical phase front structure and carrying orbital angular momentum[18], Verbeeck et. al.[5] and McMorran et. al.[6] fabricated a binary mask for the generation of off-axis vortex beams and Voloch-Bloch et. al.[7] experimented with the generation of electron Airy beams that preserve their shape and propagate in free space along curved parabolic trajectories[19]. These shaped electron beams open exciting new possibilities in electron microscopy. For example, it was shown that vortex beams can be used to characterize the magnetic state of ferromagnetic materials[5,15], whereas Airy beams can be used for realization of a new type of electron interferometer[7]. However, the experimental demonstrations of these analogies have mainly relied upon favorable natural deformations in an observed specimen[17] or, in the case of holographic projections[5–7] reconstruction in the first diffraction order using binary amplitude-based masks.

Here, we utilize widespread light-optics methods in holography and diffractive-optics to design, fabricate and experimentally measure images produced by phase-based masks using TEM. Holograms produced in this way benefit from potentially maximal energy efficiency, contrast-resolution and flexibility in their usage, marking them as the next standard in electron beam shaping.

In our experiment, the electron beams are shaped by patterning thin Silicon-Nitride (SiN) membranes using focused ion beam (FIB) milling. These membranes, ranging from 5nm to 150nm in thickness, are a popular choice due to their low scattering and mechanical robustness. Much like light waves passing through glass and acquiring a phase-shift dependent on the material's refractive index, an electron passing through a SiN membrane will similarly accumulate a phase factor directly related to the thickness of the interacting material according to[1]

$$\varphi = \frac{2\pi}{\lambda}(n-1)t = \frac{2\pi}{\lambda}\frac{eU_i}{E}\frac{E_0+E}{2E_0+E}t \quad (1)$$

where $\lambda$ is the electron's wavelength, $E_0 = m_0 c^2$ and $E = eU$ are the electron's rest and kinetic energy, respectively, where the kinetic energy is given by the acceleration voltage $U$ and the electron's charge $e$. The most interesting quantities are the thickness $t$ and the material's inner potential $U_i$, which comprises the electron-optical refractive index $n$. A relatively thin film is sufficient, e.g. for a 200 keV electron, the required thickness to generate a $\pi$-phase shift is 42nm. The scattering through this film is fairly low (several percentages) hence this can be considered a nearly pure phase plate for our purposes.

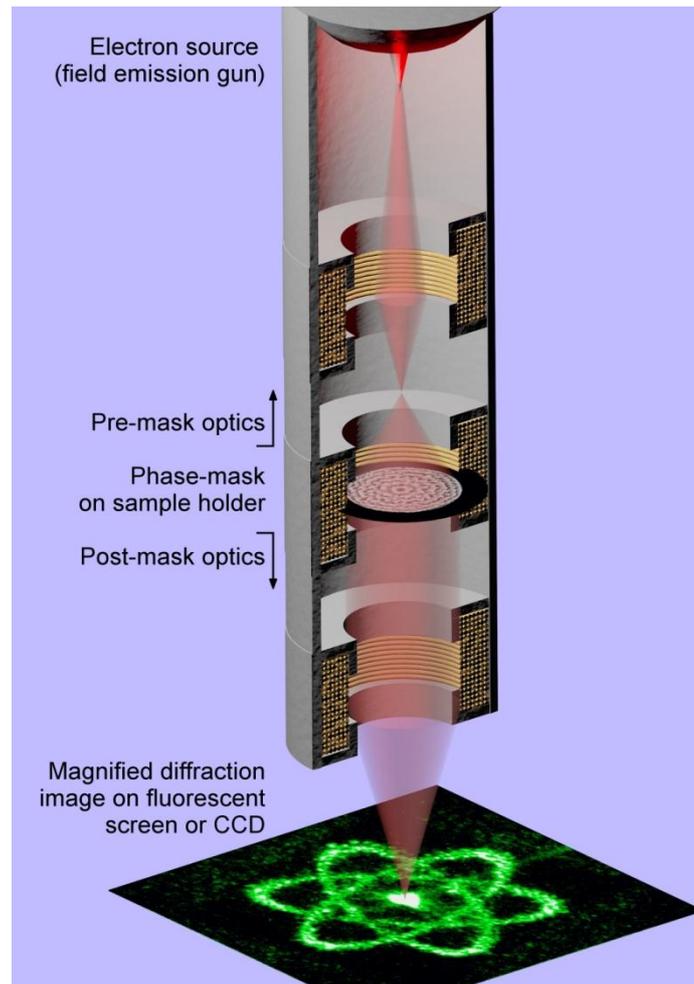

**Figure 1: Schematic of the experimental setup. The hologram (drawn here with a circular aperture) is mounted onto the sample holder which is roughly located in the objective lens system. The collimated electron beam is focused in the diffraction plane, yielding the target wavefront on the fluorescent screen or CCD.**

Computer-generated holograms (CGH) are an invaluable tool in optics. This concept conventionally allows designing a slide that stores the amplitude and phase of a wavefront, which may later be reconstructed by illuminating the hologram with a reference beam, and observing the result in the diffraction plane. CGH design may be classified as off-axis (first and higher-order) or on-axis (zero-order), the latter potentially enjoying the full conversion of energy to the desired shape.

Phase plates[4], which may be thought of as the simplest manifestations of phase-only CGHs, are generally used in electron microscopy for phase contrast[10] imaging, as first suggested by Zernike in 1942[9]. An example of such a phase plate is the Hilbert phase plate, which imparts a $\pi$-phase shift between two halves of the impinging electron beam, thus generating an approximation to the Hermite-Gauss (HG) 01 or 10 mode (which are solutions of the paraxial wave equation).

In our first experiment, described schematically in Figure 1, we studied the ability to generate such basic beams - the HG11-like as well as Laguerre-Gauss LG01-like (vortex) beams, by imparting a $\pi$ phase-shift to opposite quadrants of a disc and a continuous spiraling slope covering $2\pi$ radians, respectively. An additional periodic (Bragg) grating was fabricated; the distance between the measured diffraction orders and knowledge of the grating's period $\Lambda$ yields the effective propagation distance $L$ to the diffraction plane according to $L = \Lambda x / 2\lambda$, where $\lambda \approx 2.5\,pm$ is the electron's wavelength at 200keV, which is used as a metric for the measurements.

The evolution of the electron's wave function was measured in proximity to the diffraction plane at maximum magnification in low-angle diffraction (LAD) mode. The diffraction lens was the only lens we changed, using the Free Lens Control software, in order to traverse the propagation axis.
Figure 2 depicts these results: in (a), the beam's shape is measured passing through the membrane with no additional modulation, where its focusing and defocusing is marked by the $4\sigma$-diameter, $\sigma$ being the standard deviation of the intensity profile. In (b), the HG11-like mode shows shape invariance for an effective distance of nearly 150 meters, while the existence of a spiral phase front is evident from the enduring dark centre in (c). The effective distances are calculated from the measurements of the $\pm 1$ diffraction orders generated by the 445nm-period Bragg grating, as exemplified in (d), where $z$ denotes the relative effective distance the beam traverses in proximity to this plane. Microscopic images of the masks we used are recorded in (e-g). In the supplementary material we mathematically show that the phase plate in (f) yields a dominant mode: HG11, as clearly observed in (b).

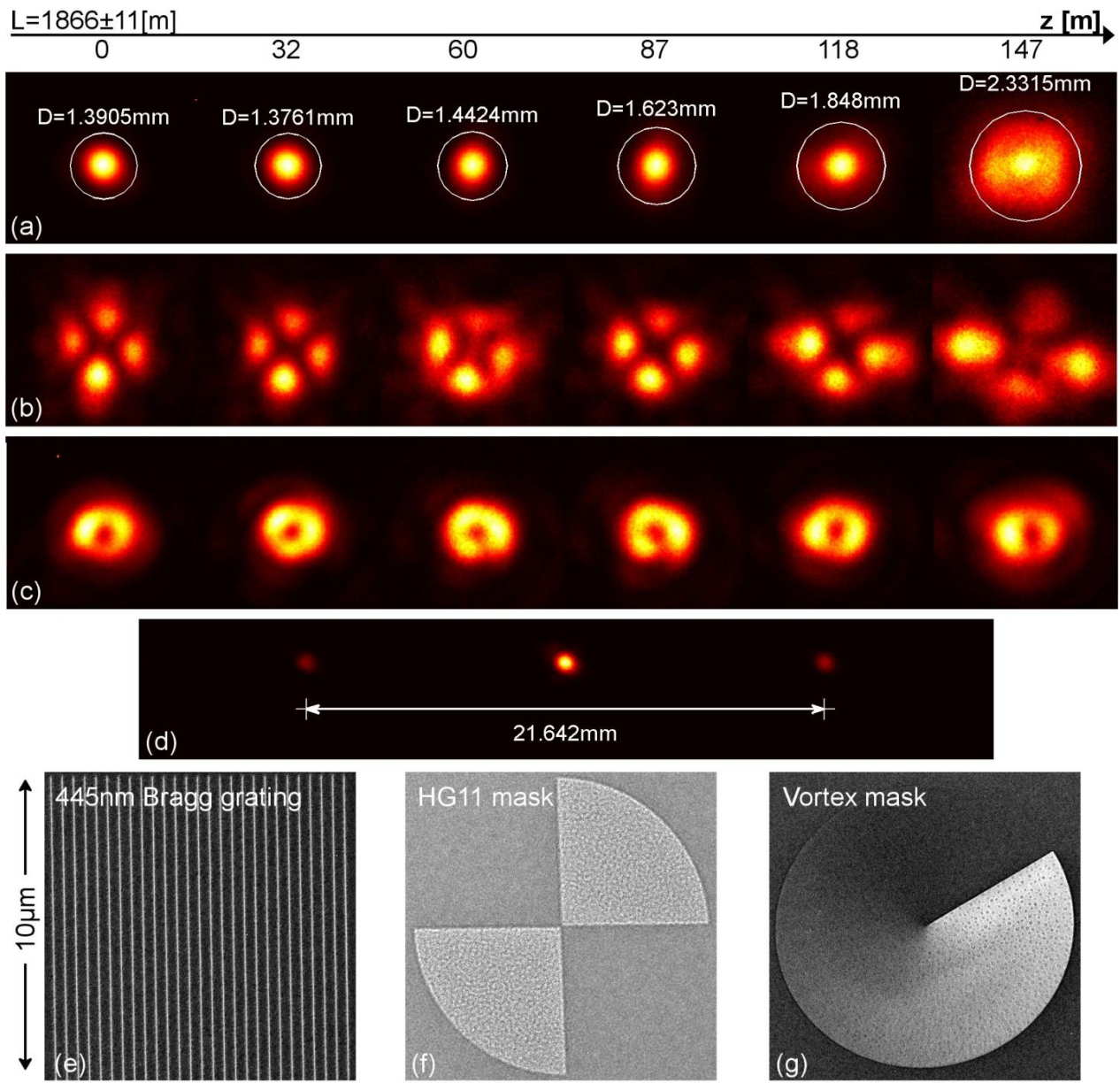

**Figure 2:** On-axis generation of electron beam free-space modes: a series of images taken at different effective distances around the diffraction plane. (a) Unmodulated beam passing through the membrane, the diameter showing the focusing and defocusing of the beam, (b) Hermite-Gauss11-like solution, (c) Laguerre-Gauss01-like solution (vortex), (d) example of a Bragg diffraction pattern used as metric, (e) Bragg grating, (f) HG11-generating mask, (g) vortex-generating mask.

It is most important to distinguish these phase-plates from previous results[5,6,8] which may be obtained with popular holographic binary schemes[20]: our beams have nearly all the energy of the impinging beam transformed to the single desired shape, on-axis. This is a major advantage with respect to the binary holograms that generate multiple diffraction orders at different angles, of which only the pattern in the first (positive or negative) diffraction order is usually required. Consequently, if the appropriate conditions are available, the vortex beam may be shrunk down to nanometer size to study, for example, electron-matter interactions such as orbital angular momentum transfer to and from the internal electron states[15]. The HG11 and similar phase-plates may be used, for instance, in conjunction with the application of phase-contrast microscopy[4] to study nearly transparent objects, e.g. biological samples[21]. The two examples given here show the ability to generate, with patterned phase masks, different electron wave functions that are solutions of the paraxial wave equation. The same technique may be utilized to generate other beams that satisfy this equation[22], e.g. higher order Hermite-Gauss and Laguerre-Gauss beams, Bessel Airy and parabolic beams, etc.

In our second experiment, we demonstrate nearly-arbitrary control of the electron's wave function by encoding two CGHs: the letters "TAU" and a model of the atom with electrons circling the nucleus. Both of these holograms were encoded on-axis using a variant of the Gerchberg-Saxton[23] iterative Fourier transform algorithm (IFTA) – an algorithm that yields a phase-mask capable of generating a desired on-axis diffraction intensity pattern. As opposed to modern fabrication technology in service of light-optics, where the fabricated feature size may easily be sub-wavelength, the significantly shorter electron's wavelength guarantees the resurgence of four fundamental difficulties in the reconstruction of on-axis holograms: speckling[24], which in our case may arise from phase discontinuities or singularities originating from the pixel's borders, multiple (two-dimensional) diffraction orders due to the finite pixel size, the appearance of a conjugate image, and the diffraction image's on-axis central spot, that is the result of constructive interference from unmodulated areas in the hologram such as spaces between pixels and fabrication errors. Despite these difficulties we can obtain high quality images of the electron beam, which are very similar to the target shapes. Measurements taken for these CGHs, presented in Figure 3, were recorded in the diffraction plane; due to the nature of the IFTA, the reconstruction of these intensity patterns is only visible near that plane. As an example, the mask (b) generated by the IFTA algorithm yields the "TAU" hologram (a). The hologram was designed to appear a small distance away from the optical axis: we may infer this by observing what seems to be the center of concentric semi-circles shifted from the centre of the mask. Indeed, in (a) the real "TAU" image is measured above the intense on-axis central spot, which is blocked to protect the CCD. In the inset, a patch of pixels is shown in relatively high magnification, boasting the FIB's capability to mill nearly-stigmatic ~60nm-diameter holes. By direct measurement of the Bragg grating's period we deduced the magnification our microscope provides, and conclude that the hologram's width in the diffraction plane is $7.6 \mu m$. This surprisingly miniscule length-scale may be reduced even further by at least two orders of magnitude, for example by placing the mask in the condenser aperture and using a stronger lens such as the microscope objective lens. However, for our specific microscope, an image created at the diffraction plane of the objective lens would already be too small to be observed on the CCD camera.

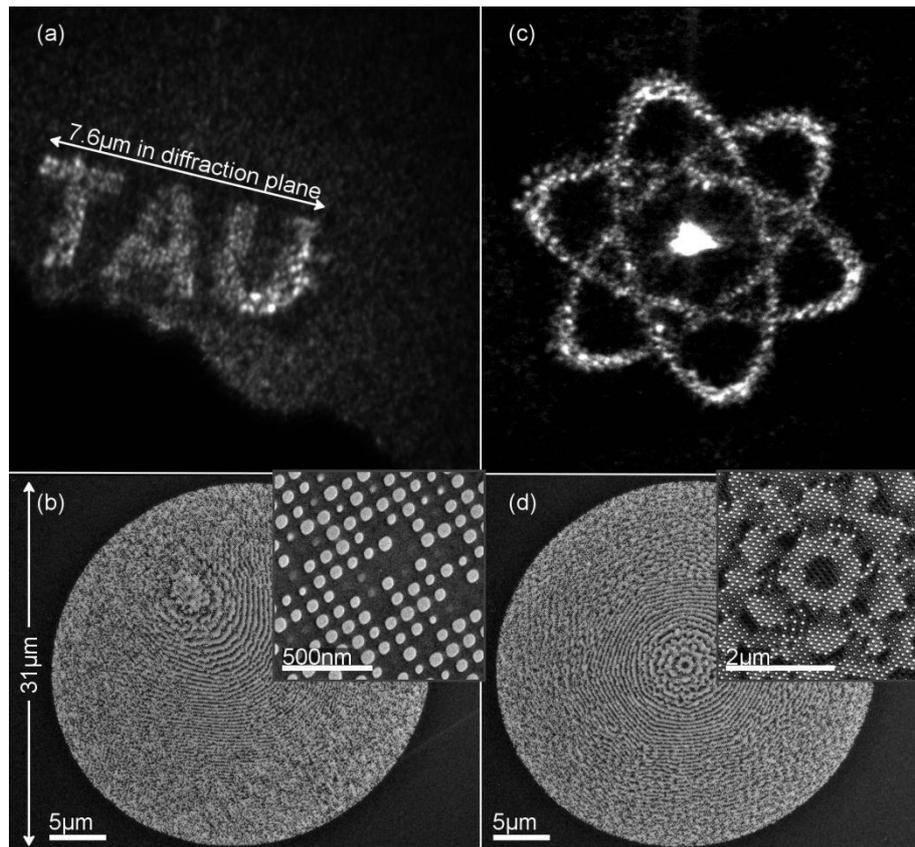

**Figure 3: On-axis holograms:** (a) "TAU" hologram produced by the mask in (b); inset: magnification showing ~60nm holes composing the pixels. (c) Electrons orbiting a nucleus hologram produced by the mask in (d); inset: magnification showing the centre of the mask. Note: contrast and brightness levels in (c) were altered for visibility.

Conversely, the mask in (d), with a magnified image provided in the inset, seems verily symmetric, a tell-tale to the hologram it yields. As may be seen in (c), the electron wave-function is molded to create a hologram featuring the classic atom model: electrons orbiting a nucleus. For the demonstration (c), taken in short exposure time, the contrast and brightness levels were altered to better show the hologram itself, which is approximately 400 times weaker than the on-axis central spot.

In this Letter we showed examples of how one can arbitrarily sculpture the electron wave-function, even within the limits of standard TEM operation. Based on the analogy between light and electron optics, we predicted and demonstrated how well-based light-optics holographic techniques may be utilized to generate shape-preserving approximations to solutions of the paraxial wave equation and arbitrary patterns using standard on-axis holographic encoding procedures. It is only recently that nano-fabrication technology has matured enough[25] to allow the engineering of holograms capable of manipulating the free electron-beam with adequate flexibility and ease. In the future, these concepts may be extended even further, beyond electron microscopy – some examples are the study of molecular matter waves[26,27]; using holograms as a basis for advanced, motionless electron-beam lithography that invalidates the need for scanning, and fast inspection by illumination with structured electron beams.

"Methods Summary"

Section 1 in the supplementary information specifies our procedure for measuring the evolution of the electron wave-function. Fabrication details are presented in Section 2. In Section 3 we derive mathematically the emergence of HG11 in the diffraction plane, after passing through the mask shown in Figure 2f.

"Acknowledgments"

The work was supported by the Israel Science Foundation, grant no. 1310/13 and the German-Israeli Project cooperation.

# Supplementary Information

1. Measuring the evolution of the electron wave-function

In order to observe the evolution of the wave-function in maximum magnification we used a 10um- or 30um-diameter objective aperture in accordance with the diameter of the hologram. This aperture is essential to obtain viable diffraction pattern. A non-modulated area of the membrane is imaged in LAD (diffraction) mode, where we found a distance of 180 meters most convenient. Making sure we are in eucentric focus (Objective lens at 6%), we focus the spot to a minimum – in this way, the beam impinging on the sample is approximately collimated. The collimation condition does not have to be strictly met to generate the holograms, but it makes possible the correct assessment of effective distances using the Bragg grating diffraction pattern. Lastly, we used the "Free lens control" software to set the magnification to the maximum possible, and bring the hologram to view by mechanically controlling the stage. Then, the evolution of the beam may be observed from image to diffraction plane and beyond, by changing the diffraction lens only.

2. Fabrication of CGHs

One side of 50- or 100nm SiN membranes was sputtered with 10nm or 5nm of either gold or titanium, respectively, to reduce charging effects while in observation under the microscope. Titanium was expected to reduce electron energy loss over gold, but in principle any sufficiently conductive material is adequate. Using an inner potential of 10V/m[1], a $\pi$-phase shift was calculated for a SiN membrane of 42nm thickness, and a number of attempts were made at different milling depths. The designs were milled into the membranes from the uncoated side, using a Raith IonLine FIB with currents ranging between 0.18pC and 0.24pC at 35kV.

All holograms required a range of $2\pi$ phase-shift and thus were fabricated on the 100nm membranes, except the HG11 phase-plate which only required a $\pi$-phase shift.

The latter plate's design is binary, whereas the vortex design consists of 36 levels ("spiraling stairs"), each fabricated as a continuous sector of growing angle. The edges of these stairs were inherently smoothed by the fabrication process.

The CGH pixels are a collection of dots, each of different depth, arranged in a rectangular lattice with 100nm period. This period was chosen after it was deduced that the FIB was capable of reliably milling the dots as 50~60nm cylinders or frustums, in order to reduce the on-axis central spot and enlarge the resolution.

3. Mathematical proof of HG11 approximation

We derive here, analytically, the emerging modes of propagation from our HG11 phase-plate. We begin by defining the normalized, 2D Hermite-Gauss function:

$$h_{m,n}(x,y) = \frac{1}{\sqrt{\pi}} \exp\left(-\frac{x^2+y^2}{2}\right) \frac{H_m(x)H_n(y)}{\sqrt{2^{m+n}m!n!}} \quad (2)$$

Where $H_m$ is the Hermite polynomial

$$H_m(x) = (-1)^m e^{x^2} \frac{d^m}{dx^m} e^{-x^2} \quad (3)$$

With $m = 0,1,2,\ldots$ We would like to expand an arbitrary 2D function in the orthonormal basis of $h_{m,n}(x,y)$ functions, $f(x,y) = \sum_{m=0}^{\infty}\sum_{n=0}^{\infty} a_{m,n} h_{m,n}(x,y)$. The coefficients $a_{m,n}$ are therefore,

$$a_{m,n} = \frac{1}{\sqrt{\pi 2^{m+n} m! n!}} \iint H_m(x) e^{-x^2/2} H_n(y) e^{-y^2/2} f(x,y) dxdy \triangleq \frac{I_{m,n}}{\sqrt{\pi 2^{m+n} m! n!}} \quad (4)$$

Our HG11-plate is defined by the sign function $f(x,y) = sign(x) sign(y)$, i.e. a four-quadrant phase mask with alternating $\pi$-shifted quadrants. We now focus on solving the double integral $I_{m,n}$ by first breaking it up into four integrals according to $f(x,y)$'s symmetry, substituting $f(x,y)$ and inverting the integral limits:

$$I_{m,n} = \left\{ \int_{x=0}^{-\infty}\int_{y=0}^{-\infty} + \int_{x=0}^{\infty}\int_{y=0}^{-\infty} + \int_{x=0}^{\infty}\int_{y=0}^{-\infty} + \int_{x=0}^{\infty}\int_{y=0}^{\infty} \right\} H_m(x) e^{-x^2/2} H_n(y) e^{-y^2/2} dxdy \quad (5)$$

Next, we change variables to account for the negative limits and apply the Hermite property $H_q(x) = (-1)^q H_q(-x)$ to write,

$$I_{m,n} = \left\{ (-1)^{m+n} + (-1)^{m+1} + (-1)^{n+1} + 1 \right\} \int_{x=0}^{\infty}\int_{y=0}^{\infty} H_m(x) e^{-x^2/2} H_n(y) e^{-y^2/2} dxdy \quad (6)$$

The expression in the curly brackets evaluates to 4 when both m and n are odd, and zero otherwise. Since the phase-mask is an anti-symmetric 2D function, this result was expected. Rewriting the integral for odd $m = 2p-1$, $n = 2q-1$ ($p, q = 1, 2, 3, ..$)

$$a_{m=2p-1, n=2q-1} = \frac{4}{\sqrt{\pi 2^{2(p+q-1)} (2p-1)!(2q-1)!}} \left[ \int_{x=0}^{\infty} H_{2p-1}(x) e^{-x^2/2} dx \right] \left[ \int_{y=0}^{\infty} H_{2q-1}(y) e^{-y^2/2} dy \right] \quad (7)$$

Evaluation of the integrals in the square brackets is not straight-forward, because of the factor $1/2$ in the exponent. Here we give an identity, which may be easily proved by differentiation and the well-known Hermite recurrence relations:

$$\int H_{2n+1}(x) e^{-x^2/2} dx = -2^{2n+1} n! e^{-x^2/2} \sum_{j=0}^{n} \frac{H_{2j}(x)}{j! 2^{2j}} + C \quad (8)$$

Applying this identity and the Hermite numbers, $H_{2j}(0) = (-1)^j (2j)!/j!$, we finally arrive to

$$a_{m=2p-1, n=2q-1} = \frac{2^{(p+q+1)}(p-1)!(q-1)!}{\sqrt{\pi(2p-1)!(2q-1)!}} \left[ \sum_{j=0}^{p-1} \frac{(-1)^j (2j)!}{(j!)^2 2^{2j}} \right] \left[ \sum_{j=0}^{q-1} \frac{(-1)^j (2j)!}{(j!)^2 2^{2j}} \right] \quad (9)$$

Which is symmetric in indices, $a_{m,n} = a_{n,m}$.

Table 1. Some values of the Hermite series coefficients $a_{m=2p-1, n=2q-1}$.

| $a_{m=2p-1,n=2q-1}$ | m=1 | 3 | 5 | 7 | 9 |
|---|---|---|---|---|---|
| n=1 | 4.5135 | 1.8426 | 2.8842 | 1.7166 | 2.4051 |
| 3 | 1.8426 | 0.7523 | 1.1775 | 0.7008 | 0.9819 |
| 5 | 2.8842 | 1.1775 | 1.8430 | 1.0969 | 1.5369 |
| 7 | 1.7166 | 0.7008 | 1.0969 | 0.6528 | 0.9147 |
| 9 | 2.4051 | 0.9819 | 1.5369 | 0.9147 | 1.2816 |

This result may be manipulated further and expressed using incomplete Beta functions. As such, the lowest coefficient $a_{1,1}$ is the largest, and it multiples the corresponding orthonormal Hermite-Gauss $h_{1,1}(x,y)$ basis functions. It is worth mentioning that these functions are bounded[28], $|h_{m,n}| \leq 0.376$.

Under our experimental conditions, the electron beam impinging on the sample may be considered of uniform illumination. Following the experiment, we must now bring the phase-shifted uniform electron beam to the diffraction plane by Fourier-transforming $f(x,y)$, and observe the resulting intensity pattern. Using the Fourier relation $F\{h_{m,n}(x,y)\} = (-i)^{m+n} h_{m,n}(k_x, k_y)$, we write:

$$|F\{f(x,y)\}|^2 = \left| \sum_{p,q=1,2,3,...} (-1)^{p+q-1} a_{m,n} h_{m,n}(k_x, k_y) \right|^2 \quad (10)$$

This final expression may be separated into positive ($p+q-1$ even) and negative ($p+q-1$ odd) sums. The result is attenuation of the lobes generated by high-order $h_{m,n}$ functions and conversely an amplification of the $h_{1,1}$ lobes, as observed in Figure 4.

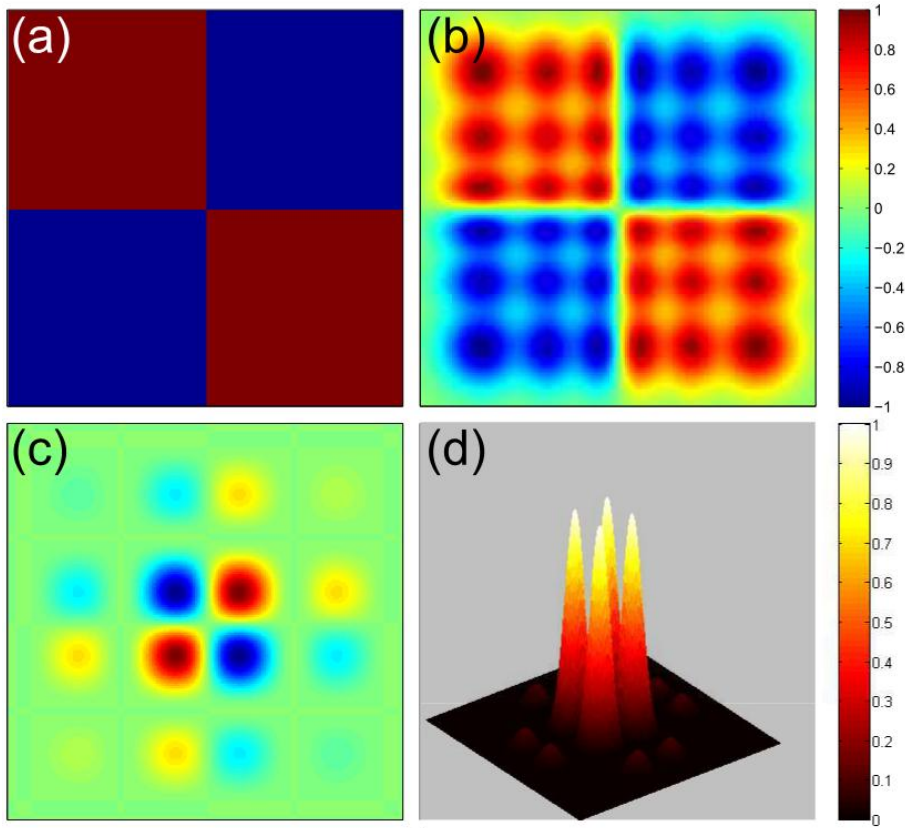

**Figure 4: Visualization of the mathematical derivation.** (a) The function $f(x,y) = sign(x)sign(y)$, a four-quadrant mask with alternating +/-1 quadrants. (b) a truncated Hermite-Gauss expansion of $f(x,y)$ with the 25 $m,n$ terms from Table 1. (c) Real part of the Fourier transform of (b), and (d) absolute square of the latter, which is proportional to the resulting intensity pattern. All values are normalized but linear in scale.

Coupled with the fact that the measured signal is an intensity pattern (i.e. the absolute signal squared), all of these arguments lead to the domination of the HG11 mode in the diffraction plane, as measured in the experiment. Furthermore, it is important to note that the circular objective aperture may mathematically be introduced as a convolution with a Bessel function, the main effect of which is the truncation of the higher-order Hermite-Gauss functions in the expansion.

It is interesting to note that though we measure the HG11-like pattern in close proximity to the focal point, away from it each term in the series, a Hermite-Gauss mode, propagates according to the wave equation with different normalized beam parameter product, $M^2$ (i.e. they diffract differently[29]), until the beam completely separates. Since the lower order modes (HG00, HG01 and HG10) are not present, the HG11 mode, with $M^2 = 3$, is the slowest to diffract.